\pdfoutput=1

\documentclass{ws-jai}
\usepackage[flushleft]{threeparttable}
\usepackage{epstopdf}
\usepackage{longtable}
\usepackage{tabularx}

\begin{document}

\catchline{}{}{}{}{} 

\markboth{Tillman, Ellingson, and Brendler}{Practical Limits in the Sensitivity-Linearity Trade off for Radio Telescope Front Ends in the HF and VHF-low Bands}

\title{Practical Limits in the Sensitivity-Linearity Trade off for Radio Telescope Front Ends in the HF and VHF-low Bands}

\author{R.H. Tillman$^{\dagger*}$, S.W. Ellingson$^\dagger$, and J. Brendler$^\dagger$}

\address{
$^\dagger$Bradley Department of Electrical and Computer Engineering, Virginia Tech, Blacksburg, VA 24060, USA, hanktillman@vt.edu
}

\maketitle

\corres{$^*$hanktillman@vt.edu.}

\begin{history}
\received{(to be inserted by publisher)};
\revised{(to be inserted by publisher)};
\accepted{(to be inserted by publisher)};
\end{history}

\begin{abstract}
Radio telescope front ends must have simultaneously low noise and sufficiently-high linearity to accommodate interfering signals.
Typically these are opposing design goals.
For modern radio telescopes operating in the HF (3--30~MHz) and VHF-low (30--88~MHz) bands, the problem is more nuanced in that front end noise temperature may be a relatively small component of the system temperature, and increased linearity may be required due to the particular interference problems associated with this spectrum.   
In this paper we present an analysis of the sensitivity-linearity trade off at these frequencies, applicable to existing commercially-available monolithic microwave integrated circuit (MMIC) amplifiers in single-ended, differential, and parallelized configurations. 
This analysis and associated findings should be useful in the design and upgrade of front ends for low frequency radio telescopes.
The analysis is demonstrated explicitly for one of the better-performing amplifiers encountered in this study, the Mini-Circuits PGA-103, and is confirmed by hardware measurements.
We also present a design based on the Mini-Circuits HELA-10 amplifier, which is better-suited for applications where linearity is a primary concern. 
\end{abstract}


\section{Introduction}
Modern low frequency ($<300$~MHz) radio telescopes consist of dozens to thousands of low gain antennas, the outputs of which are typically digitized so beamforming and imaging may be performed in the digital domain.
In these telescopes, each individual antenna is integrated with a front end, which typically performs amplification, filtering, impedance transformation, and differential to single-ended (i.e. balanced to unbalanced) conversion. 
The front end must be sensitive to astrophysical signals, but must also be sufficiently linear to be robust to external interference.
Thus there exists a trade off between sensitivity and linearity.

In this paper we present front end design methodologies which consider sensitivity and linearity jointly.
System-level sensitivity requirements place well-known constraints on the receiver noise temperature $T_R$, leading to typical values in the range of a few hundred kelvin~\cite{Ellingson05}.
Linearity requirements are less well established.
Table~\ref{tab:FEEComp} compares the front ends employed in some modern low frequency radio telescopes: the first station of the Long Wavelength Array (LWA1)~\cite{Ellingson13}, the Murchison Widefield Array (MWA)~\cite{Tingay2013}, the Low Frequency Array (LOFAR) Low-Band~\cite{vanHaarlem_2013}, the Expanded Very Large Array (EVLA) ``4-Band'' system~\cite{Kassim2007}, and the Eight-meter wavelength Transient Array (ETA)~\cite{Deshpande2009}.	

To quantify the sensitivity-linearity trade off, we performed a survey of commercially-available monolithic microwave integrated circuit (MMIC) amplifiers that might be considered for use in radio telescope front ends.
From this survey we identified two amplifiers: the Mini-Circuits (MC) PGA-103, a standout in $T_R$, and the MC HELA-10, a standout in linearity.
We designed amplifiers using PGA-103s in (1) a balanced configuration and in (2) parallel configuration of balanced amplifiers in order to improve the linearity of the single-ended PGA-103.
A front end using two HELA-10 amplifiers with input filtering was designed to demonstrate a front end with the highest reported linearity in a modern radio telescope.
The relevant performance metrics for these new front end designs are also included in Table~\ref{tab:FEEComp}.

The remainder of this paper is organized as follows.
Section~\ref{sec:NoiseTempReqs} summarizes the sensitivity requirements for front ends in modern low frequency radio telescopes.
In Section~\ref{sec:Survey} we present a survey of commercially-available MMIC amplifiers.
Section~\ref{sec:AmpModel} quantifies the performance of amplifiers consisting of a parallel configuration of balanced amplifiers, in terms of the linearity-sensitivity trade off.
Sections~\ref{sec:PGA}~and~\ref{sec:HELA} present the PGA-103 (high sensitivity) and HELA-10 (high linearity) front ends, respectively.
Section~\ref{sec:Demo} presents the results of a field test demonstration using the front ends presented in Section~\ref{sec:DesignExamples}.
Our conclusions are summarized in Section~\ref{sec:conc}.

	
\begin{rotatetable}
\begin{tabular}{l r r r r l}
\hline
Front End & \multicolumn{1}{l}{$\nu$} & \multicolumn{1}{l}{$T_R$} & \multicolumn{1}{l}{Input $P_{1dB}$} & \multicolumn{1}{l}{DC Power} & \multicolumn{1}{l}{Reference} \\ 
 & \multicolumn{1}{c}{(MHz)} & \multicolumn{1}{r}{(K)} & \multicolumn{1}{c}{(dBm)} & \\ \hline \hline
EVLA ``4-Band''$^\dagger$ & 50-86 & 710 & $-$13 at ~~74 MHz & 2000~mA @ 17.5 V & (Harden, P. (NRAO), Personal Comm.) \\
ETA & 27-49 & 300 & $-$3 at ~~38~MHz & 160 mA @ 12~~ V & \cite{ESP07} \\
LWA1 & 10-88 & 255 & $-$18 at ~~74~MHz & 260 mA @ 15~~ V & \cite{HICKS_PASP_2012} \\
LOFAR~Low-Band$^\ddagger$ & 10-80 & 180 & $-$12 at 100~MHz & 120 mA @ ~3~~ V & \cite{Wijnholds_2011} \\
MWA$^\ddagger$ & 80-300 & 21 & $-12$ at 100~MHz & 120 mA @ ~3~~ V & \cite{Lonsdale2009} \\
\hline
\multicolumn{5}{c}{This Work} \\ \hline
HELA-10 & 30-80 & 700 & $+$11~~ at 50~MHz & 800~mA @ 18~~~V & Section \ref{sec:HELA} \\
PGA-103 - Parallel & 30-80 & 131 & $+8.3$ at 50~MHz & 90~mA @ 12~~ V & Section \ref{sec:PGA} \\
PGA-103 - Balanced & 30-80 & 112 & $+3.3$ at 50~MHz & 180~mA @ 12~~ V & Section \ref{sec:PGA} \\
PGA-103 - Single-Ended & 30-80 & 51 & $-0.5$ at 50~MHz & 360 mA @ 12~~ V & Section \ref{sec:PGA} \\
\hline
\end{tabular}
\caption{Comparison of the fronts ends used in some modern low frequency radio telescopes.
	See text for definitions and nomenclature.
	$^\dagger$The reported $T_R$ includes 2~dB cable loss between antenna feed and front end.
	$^\ddagger$The measured $P_{1dB}$ for the front ends of LOFAR and MWA are not known to us; this value is for the transistor used in the respective front ends under bias conditions recommended in the datasheet.}
\label{tab:FEEComp}
\end{rotatetable}

\section{Noise Temperature Requirements}
\label{sec:NoiseTempReqs}

As in any other frequency regime, the sensitivity of a low frequency radio telescope is determined primarily by collecting area and noise.  
However the extraordinarily high sky brightness temperature at  low frequencies leads to somewhat different considerations in design criteria and subsequent characterization of instruments.  
  
\begin{figure}
\centerline{\includegraphics[width=3in]{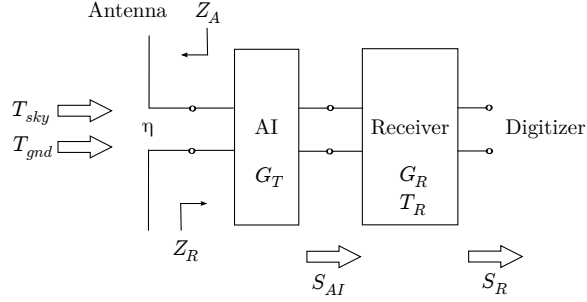}}
\caption{Model for analysis of noise in low-frequency instruments.}
\label{lfa_fNoiseModel}
\end{figure}

A model for characterizing the sensitivity-limiting noise delivered to the digitizer in a low frequency instrument is shown in Fig.~\ref{lfa_fNoiseModel}.
For simplicity this particular model assumes that each element is individually digitized, but the principles remain relevant when analog combining is employed. 
Noise captured by the antenna is conveniently characterized in terms of antenna temperature $T_A$.
For the purposes of this paper $T_A$ is defined as the power spectral density (PSD; e.g., W~Hz$^{-1}$) delivered to a load which is conjugate-matched to the antenna impedance $Z_A = R_A + jX_A$, divided by Boltzmann's constant $k=1.38 \times 10^{-23}$~J/K.

As will be explained later, antennas used in low frequency arrays are often poorly-matched to receivers.  
The effect of impedance mismatch may be taken into account by defining a two-port circuit, identified as an ``antenna interface'' (AI) in Fig.~\ref{lfa_fNoiseModel}, which connects the antenna to the rest of the receiver.  
The AI is assumed to be perfectly-matched to the receiver over the bandwidth of interest, and exhibits input impedance $Z_R = R_R + j X_R$ which may or may not vary significantly with frequency.  
In implementation terms, the AI may be any combination of impedance transforming devices, baluns, and filters; or the AI may simply a ``null'' two-port representing direct connection between the antenna and receiver.
In any event, the AI may be characterized in terms of transducer power gain $G_T$, which in the present problem may be defined as the ratio of the PSD $S_{AI}$ delivered to the receiver, to the PSD $kT_A$ ``available'' from the antenna.  
Neglecting for the moment any noise that might originate from the AI (i.e. assuming the AI is lossless), 
\begin{equation}
S_{AI} = k T_A G_T 
\end{equation}   
where 
\begin{equation}
G_T = \frac{ 4 R_A R_R }{ \left| Z_A + Z_R \right|^2 }
\end{equation}
as may be verified using elementary circuit analysis.
$G_T$ may alternatively be expressed in terms of the ``power wave reflection coefficient'' $\widetilde{\Gamma}_{A}$\footnote{
Note an easy-to-make mistake: Eq.~(\ref{eqn:G_T_IME}) is invalid when the voltage reflection coefficient $\Gamma_A$ (no tilde) is used in lieu of $\widetilde{\Gamma}_{A}$: i.e.,
\begin{equation*}
	G_T \neq 1 - \left| \Gamma_{A} \right| ^2 \quad , \text{where} \quad \Gamma_{A} = \frac{Z_A - Z_R}{Z_A + Z_R}
\end{equation*}
unless either $Z_A$ or $Z_R$ is purely real valued.
}
\begin{equation}
	G_T = 1 - \left| \widetilde{\Gamma}_{A} \right| ^2 \quad , \text{where} \quad \widetilde{\Gamma}_{A} = \frac{Z_A - Z_R^*}{Z_A + Z_R}
	\label{eqn:G_T_IME}
\end{equation}
\cite{TillmanEllingson2015b}. 
The PSD subsequently delivered to the digitizer is
\begin{equation}
S_R = kT_A G_T G_R + kT_R G_R
\label{eqn:Sr}
\end{equation}
where $G_R$ is the gain of the receiver and the second term accounts for the noise contribution from the receiver, expressed as the input-referred equivalent noise temperature $T_R$.

Front ends in low frequency radio telescopes are not typically noise-matched to the antenna.
Antenna impedance in these instruments varies dramatically, so the front end input impedance is typically not equal to the value that would yield optimal noise figure.
As a consequence, the minimum possible $T_R$ is not obtained.
However, as long as the realized $T_R / G_T$ is sufficiently below $T_A$, improved noise matching will have only a minor impact on sensitivity (more on this below; see paragraph including Eq.~\ref{eqn:StopCrit}).
Thus, Eq.~(\ref{eqn:Sr}) is exact only if $T_R$ is determined for the \textit{in situ} input impedance, as opposed to using the value associated with the vendor-specified impedance; but the effect on the results presented here is expected to be small since the mismatch with the antenna impedance remains large.

$T_A$ is comprised of naturally-occurring contributions $T_{sky}$ from the sky and $T_{gnd}$ from the ground, plus noise associated with interference.
Interfering noise originates from human activity and from natural events including lightning and solar bursts~\cite{ITU_2015}.  
In contrast to deliberate man-made signals, man-made noise below 300~MHz is normally negligible compared to naturally-occurring noise in the rural locations where radio telescopes are typically built.
Naturally-occurring interference is intermittent and negligible most of the time, so the resulting antenna temperature can be expressed as
\begin{equation}
T_A = \eta \left( T_{sky} + T_{gnd} \right)
\label{eqn:TA}
\end{equation} 
where $\eta$ is ground loss efficiency.  This efficiency accounts for near-field coupling between the antenna and the ground that results in loss in the ground manifesting as loss in the antenna itself.  Typical values of $\eta$ range from about $0.5$ (typical at 38~MHz above untreated earth without a conducting ground screen \cite{ESP07}) to very close to 1 (typical above 100~MHz or with a sufficiently large ground screen, as in the demonstration in Section~\ref{sec:Demo}).
The value of $T_{gnd}$ is significantly affected by the ground screen.
For a perfectly-conducting infinite ground screen, the apparent brightness of the ground is about equal to the apparent brightness of the sky, due to reflection. For no ground screen $T_{gnd} \sim $150~K, i.e., $\sim$300~K corresponding to the physical temperature of the ground divided by 2 since roughly half of the antenna pattern intersects the ground.
For the small ($\mathcal{O}[\lambda^2]$ in area) ground screens typically used in modern instruments, the actual value of $T_{gnd}$ is somewhere between these two extremes (for example, we obtain a value of 390~K at 45~MHz for the field experiment conducted in Section~\ref{sec:Demo}).
$T_{sky}$ in this frequency regime is dominated by the very bright Galactic synchrotron background.
The brightness temperature of this background is somewhat brighter in the Galactic plane and somewhat dimmer in the Galactic polar regions, however the total contribution to the antenna temperature of a low-gain antenna isolated from ground loss is well-approximated as 
\begin{equation}
T_{sky} \approx \left(9120~\mbox{K}\right) \left( \frac{\nu}{39~\mbox{MHz}} \right)^{-2.55}
\label{eqn:lfa_eTsky}
\end{equation}
where $\nu$ is frequency. 
(This expression is obtained by a fit to the model described in Appendix~I of \cite{Ellingson05}.)
This is shown in Fig.~\ref{lfa_fTant}.
Note that a diurnal variation of about $\pm20$\% is present in this contribution, with the maximum corresponding to the transit of the Galactic center. 
%
\begin{figure}
\centerline{\includegraphics[width=\textwidth]{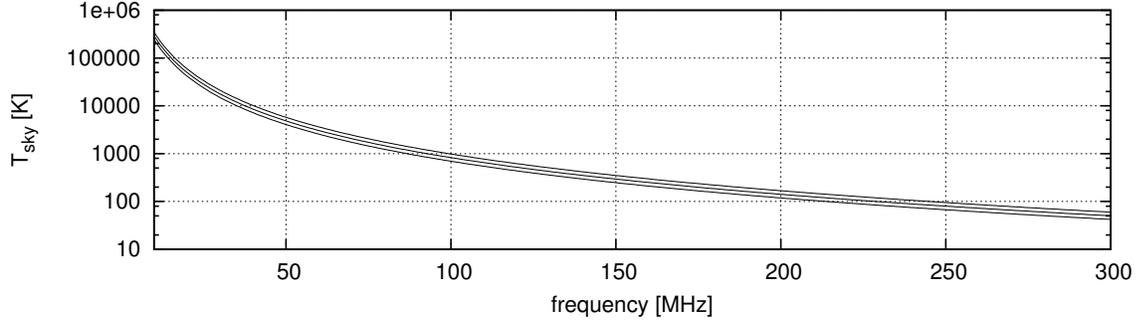}}
\caption{Contribution of the Galactic synchrotron background to the antenna temperature of a dipole-type antenna isolated from ground loss.  The center curve is the daily mean value whereas the upper and lower curves represent the typical limits due to diurnal variation at about 35$^\circ$N latitude.}
\label{lfa_fTant}
\end{figure}

\begin{figure}
\begin{center}
\begin{tabular}{cc}
\includegraphics[width=2.40in]{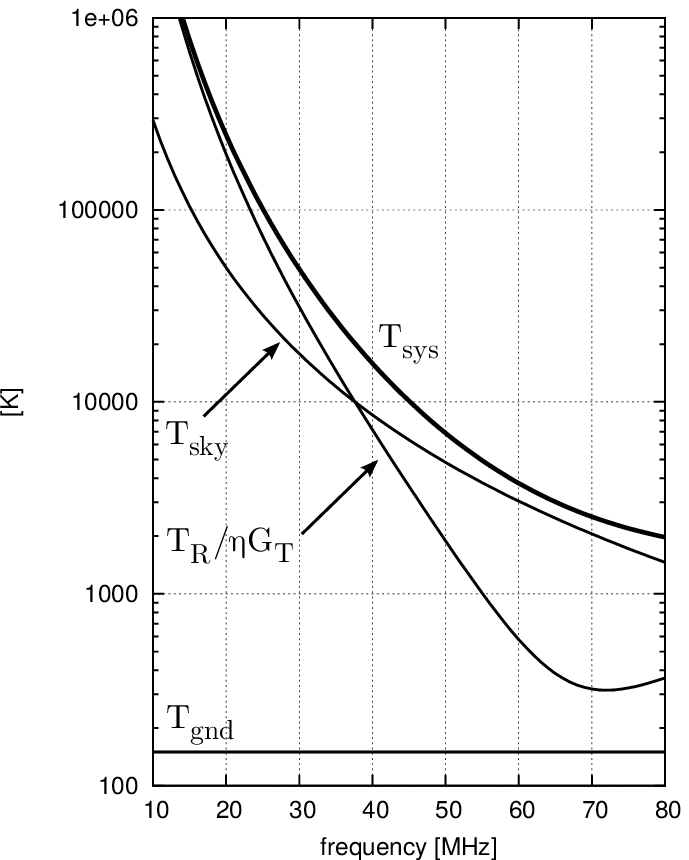} & 
\includegraphics[width=2.16in]{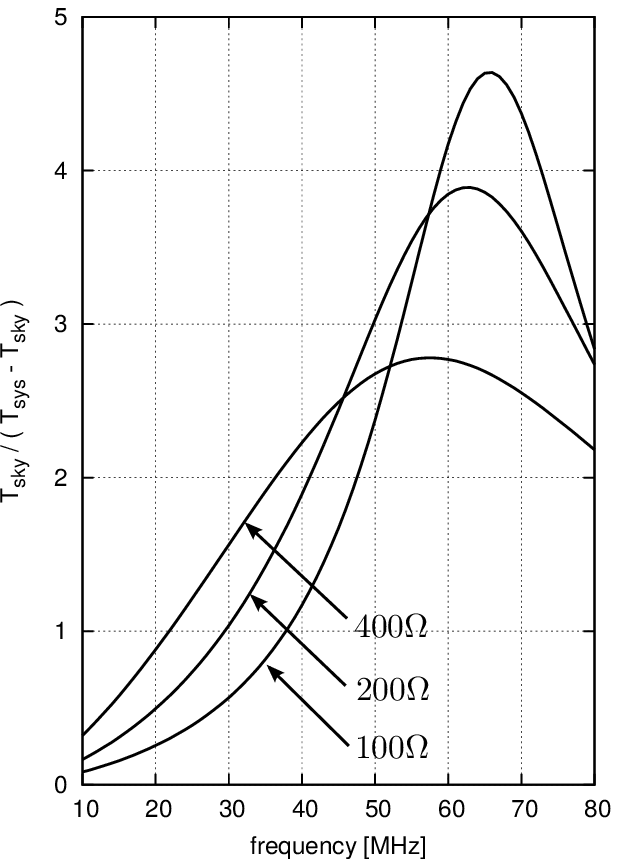} \\
\end{tabular}
\end{center}
\caption{{\it Left:} Contributions to the apparent system temperature ($Z_R=100\Omega$). {\it Right:} $T_{sky}$ relative to other sources of noise for various receiver input impedances $Z_R$. See text for scenario details.}
\label{lfa_fNoiseTerms}
\end{figure}

Combining results, the PSD delivered to the digitizer is
\begin{equation}
S_R = \eta k T_{sky} G_T G_R + \eta k T_{gnd} G_T G_R + k T_R G_R
\label{lfa_eSR}
\end{equation}
To convey a sense of magnitudes, Fig.~\ref{lfa_fNoiseTerms} shows these contributions individually and together for a simple dipole 2-m long by 2~cm in diameter.
The antenna is half-wavelength resonant at $\approx70$~MHz. 
In this result, $Z_A$ is obtained using an equivalent circuit model~\cite{TTG93}, 
$Z_R=100~\Omega$, 
$T_{sky}$ is from Eq.~(\ref{eqn:lfa_eTsky}), 
$T_{gnd}=150$~K, 
$\eta=1$, 
and 
$T_R = 300$~K. 
The contributions are shown in the form of input-referred equivalent noise temperature, which is essentially Eq.~(\ref{lfa_eSR}) divided by $\eta k G_T G_R$ so that the sum may be interpreted as system temperature; i.e.
\begin{equation}
	T_{sys} = \frac{S_R}{\eta k G_T G_R} = T_{sky} + T_{gnd} + \frac{T_R}{\eta G_T}
	\label{eqn:T_sys}
\end{equation}
  
It is apparent that the noise delivered to the digitizer is dominated by $T_{sky}$ over a large fraction of the frequency range considered.  
This is desirable since the Galactic noise contribution is irreducible and so any further improvements cannot significantly improve sensitivity.  
From this perspective the following ``stopping criterion'' for design effort may be obtained from Eq.~(\ref{eqn:T_sys}):
\begin{equation}
	\frac{T_R}{G_T} \ll \eta \left( T_{sky} + T_{gnd} \right)
	\label{eqn:StopCrit}
\end{equation}
In other words, peak sensitivity is essentially optimized by making $T_R$ sufficiently small relative to the AI impedance mismatch (represented by $G_T$), such that further optimization of $T_R$ or $G_T$ then has little benefit.

While peak sensitivity is limited by $T_R/G_T$, it is apparent from Fig.~\ref{lfa_fNoiseTerms} that the {\it bandwidth} over which this ``optimal'' sensitivity is achieved may be improved by either further reducing $T_R$, or by expanding the bandwidth associated with $G_T$.  Further reduction in $T_R$ is normally not attractive, since this typically entails a corresponding reduction in linearity.   Alternatively, the antenna may be ``broadbanded'' such that $G_T$ varies less over the bandwidth of interest, or -- less obvious as a solution -- $Z_R$ might be increased.  The broadbanding strategy leads to ``fat dipoles''; prime examples being the antenna elements used by LWA1, MWA, and the LOFAR high-band array. 
In the second strategy, $Z_R$ is increased by impedance transformation within the AI~\cite{Ellingson05}.
This is demonstrated in Fig.~\ref{lfa_fNoiseTerms}, where it is apparent that increasing $Z_R$ decreases peak sensitivity but increases the bandwidth over which a specified minimum level of sensitivity can be achieved.
Note that the vertical axis of this plot can also be interpreted as a measure of the integration time required to achieve a specified signal-to-noise ratio~\cite{Erickson2005}.



\section{Linearity-Sensitivity Trade off for Existing Commercially-Available MMIC Amplifiers}
\label{sec:Survey}

There is a myriad of commercially available MMIC amplifiers that would appear to be suitable for modern low frequency radio telescopes (see Table~\ref{tab:FEEComp}).
A process for selection of an ``optimal'' amplifier is not obvious. 
However, amplifiers exhibit a trade off between linearity and sensitivity, which may be used to identify exceptional amplifiers.
To quantify the linearity-sensitivity trade off in commercially-available MMIC amplifiers, we performed a market survey of  271 devices from Mini-Circuits (MC), TriQuint, Analog Devices, Hittite (recently acquired by Analog Devices), and Avago Technologies, all having noise figure\footnote{$\text{NF} = 10 \log_{10} \left( T_R/(290~\text{K}) + 1 \right) $.} (NF)$<$9~dB ($\sim 2000$~K) and advertised as suitable for operation at 50~MHz. 
The raw data collected for this survey is freely available from the authors.
	
\begin{figure}
\begin{center}
\begin{tabular}{cc}
\includegraphics[width=3in]{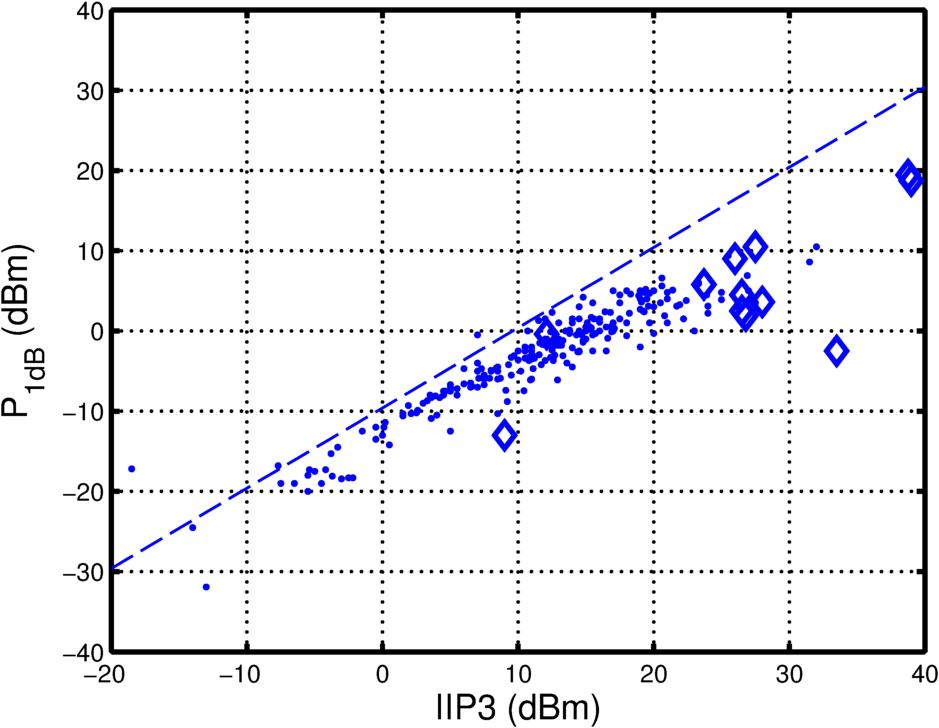} & 
\includegraphics[width=3in]{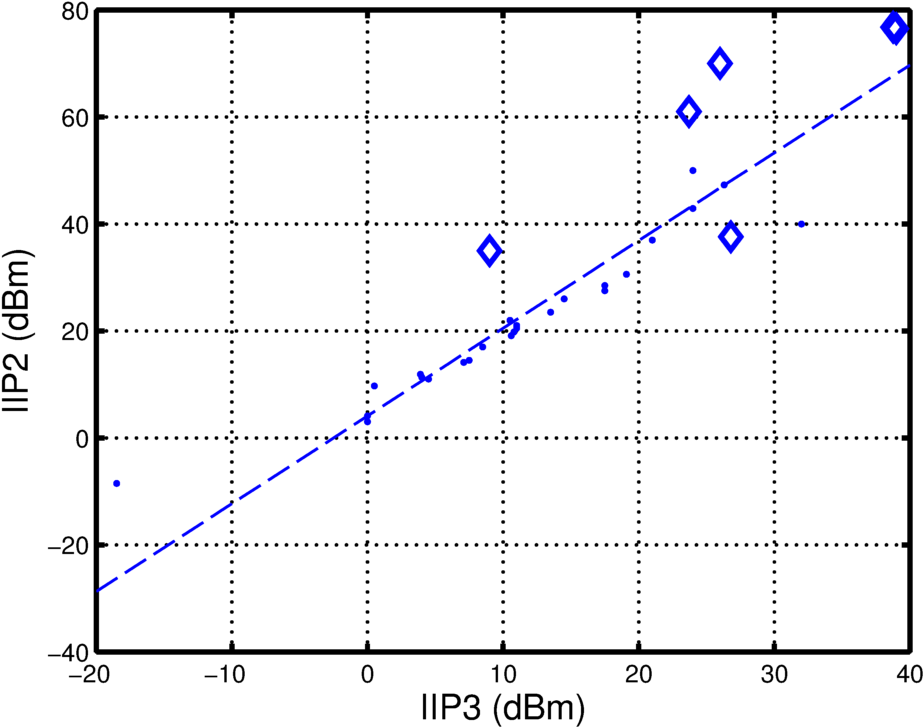} \\
\end{tabular}
\end{center}
\caption{ {\it Left:} Relationship between $P_{1dB}$ and IIP3 for amplifiers in the survey.
	The dashed line shows the theoretical relationship $P_{1dB} = \text{IIP3} - 9.6$~dB for systems exhibiting third-order memoryless non-linearity~(see e.g. \cite{Razavi12}).
	{\it Right:} IIP2 and IIP3  for amplifiers in this survey for which both are reported.
	The dashed line shows the linear least squares fit. 
	Differential amplifiers are identified by a diamond marker in both plots.}
\label{fig:LinearityMetrics}
\end{figure}

The linearity-sensitivity trade off was assessed by comparing the input-referred third-order intercept point (IIP3) to $T_R$.
IIP3 is defined as the power input to an amplifier at which the amplitude of the output third-order intermodulation products (i.e., the cross-frequency terms output given a two-tone input signal) is equal to the amplitude of the fundamental tone.
Two other relevant linearity metrics are the input-referenced second-order intercept point (IIP2), defined as the power of a tone input to an amplifier at which the amplitude of the output second-order intermodulation product (i.e. second harmonic) is equal to the amplitude of the fundamental tone; and the input 1~dB compression point ($P_{1dB}$), defined as the power input to a system at which the system's gain is reduced by 1~dB from the expected (nominally linear) gain.
IIP2 is an important consideration in receivers having high fractional bandwidth, since the second harmonic of an interfering signal is more likely to fall within the receiver's pass band.
However IIP2 is typically not reported by manufacturers.
On the other hand, IIP2 and $P_{1dB}$ were found to be highly correlated to IIP3 for the amplifiers in this survey, as shown in Fig.~\ref{fig:LinearityMetrics}.
Thus, IIP3 serves as a reasonable, if imperfect, metric for overall linearity.
  
\begin{figure}
	\centering{\includegraphics[width=\textwidth]{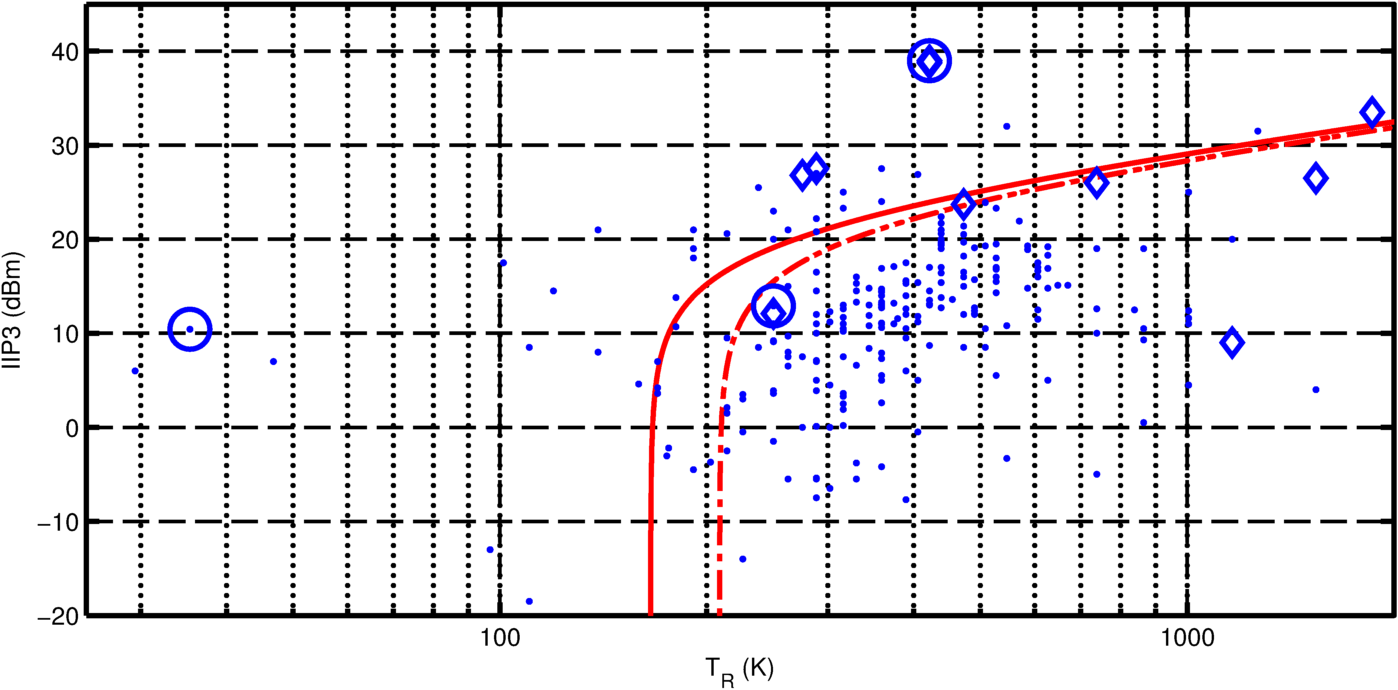}}
	\caption{Linearity versus sensitivity for all 271 amplifiers considered in the survey.
		Points represent datasheet values for individual amplifiers for the recommended bias point and 50~MHz.
		Differential amplifiers are identified by diamond markers.
		The dot-dash and solid lines are a bound the majority and outlying populations, respectively.}
	\label{fig:Survey}
\end{figure}

   
The results of the survey are summarized in Fig.~\ref{fig:Survey}.
The curves in Fig.~\ref{fig:Survey} are a fit by eye to the equation 
\begin{equation}
	\text{IIP3} = a \text{F} + b
\end{equation}
where F is the noise factor (NF in linear units), IIP3 is in Watts, and the constants $a$ and $b$ are selected to obtain curves that approximately bound the performance of the populations of lower- and higher-performing amplifiers.
The two curves are shown in Fig.~\ref{fig:Survey} to illustrate the gap between the majority of amplifiers ($a=250$, $b=-430$) and outlying amplifiers $(a=280$, $b=-440$). 
The MC PGA-103 (used in Section~\ref{sec:PGA}), HELA-10 (used in Section~\ref{sec:HELA}), and GALI-74 (used in LWA1) are circled at (35~K, 10.4~dBm), (422~K, 39~dBm), and (250~K, 12.9~dBm), respectively.

\section{Sensitivity and Linearity for Single-ended, Differential, and Parallelized Amplifier Configurations}
\label{sec:AmpModel}

A front end consisting of multiple MMIC amplifiers in balanced and/or parallel configurations will have increased IIP3 and increased NF, resulting in a shift up and to the right of the associated single-amplifier front end in Fig.~\ref{fig:Survey}.  
This shift in performance can be predicted, as will be described below. 

\begin{figure}
	\centering
	\includegraphics[scale=1]{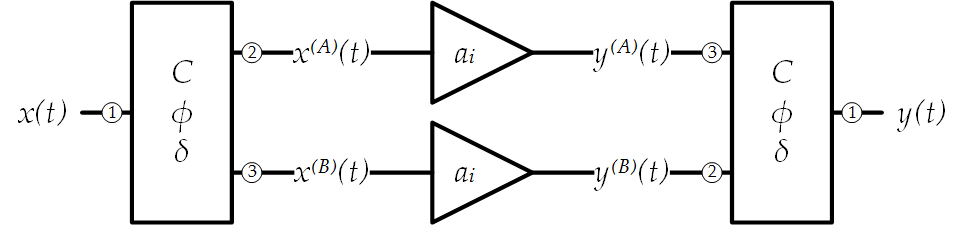}
	\caption{Balanced amplifier topology.
		See Section~\ref{sec:AmpModel} for nomenclature.
		The phase shift $\phi$ is nominally $180^\circ$.}
	\label{fig:QuadComb}
\end{figure}

It is also well known that linearity can be improved by distributing the input power over two amplifiers arranged in a balanced configuration as shown in Fig.~\ref{fig:QuadComb}; see e.g. \cite{ENGELBRECHT_ProcIEEE_1965}.
This can be achieved by distributing power to the amplifier inputs using a passive hybrid with outputs $180^{\circ}$ out of phase, and then similarly combining from the amplifier outputs. 
Further improvement in linearity may be achieved by arranging single-ended or balanced amplifiers in parallel, for example using hybrids for power splitting and combining at the input and output, respectively.
To the best of our knowledge an analysis of $T_R$ and IIP3 jointly for amplifiers in this class of topologies has not been previously reported.

\subsection{Linearity}
 
Consider a single amplifier whose voltage (or current) transfer function is assumed to be well-modeled as a third-order power series $y(t) = a_1 x(t) + a_2 x^2(t) + a_3 x^3(t)$.
IIP2 and IIP3 can be determined by applying a two tone signal $x(t) = A [ \cos \omega_1 t + \cos \omega_2 t ]$ to the amplifier.
The output signal consists of sinusoidal components representing the desired tones with gain compression, a DC component, second and third harmonics of $\omega_1$ and $\omega_2$, second-order intermodulation products~(IM2) at $\omega_2 \pm \omega_1$, and third-order intermodulation products~(IM3) at $2 \omega_2 \pm \omega_1$ and $2 \omega_1 \pm \omega_2$.
In this case, IIP2 and IIP3 may be written in voltage or current units as
\begin{equation}
	A_{IIP2} = \left| \frac{a_1}{a_2} \right|.	\quad	A_{IIP3} = \sqrt{\frac{4}{3} \left| \frac{a_1}{a_3} \right| }
\label{eqn:N0}
\end{equation}
see e.g.~\cite{Razavi12}.

Next, consider a single pair of amplifiers in a balanced configuration, as shown in Fig.~\ref{fig:QuadComb}.
The hybrids are assumed to be impedance-matched reciprocal, frequency-independent devices.
Further, the hybrids are characterized by a coupling factor $C$, which is the ratio of power  into port 1 to power out of port~2 or port~3; phase shift $\phi$, nominally $180^\circ$; and phase error $\delta$.
Any loss in the splitter is included in $C$.
As derived in Appendix~A, the IIP2 and IIP3 for the single pair of balanced amplifiers are
\begin{equation}
	A_{IIP2}^{(1)} \cong 2 \frac{\sqrt{C}}{\delta} A_{IIP2}^{(0)} \quad , \quad A_{IIP3}^{(1)} \cong \sqrt{C} A_{IIP3}^{(0)}
	\label{eqn:BallancedIIP}
\end{equation}
where $A_{IIP2}^{(0)}$ and $A_{IIP2}^{(1)}$ are the IIP2 for the single amplifier and balanced amplifier, respectively, and similarly for $A_{IIP3}^{(0)}$ and $A_{IIP3}^{(1)}$.
As expected, one finds that the intercept points are improved in proportion to $\sqrt{C}$ (i.e. proportional to $C$ in power units), and that IIP2 is impaired by $\delta \neq 0$.

\begin{figure}
	\centering{\includegraphics[scale=0.75]{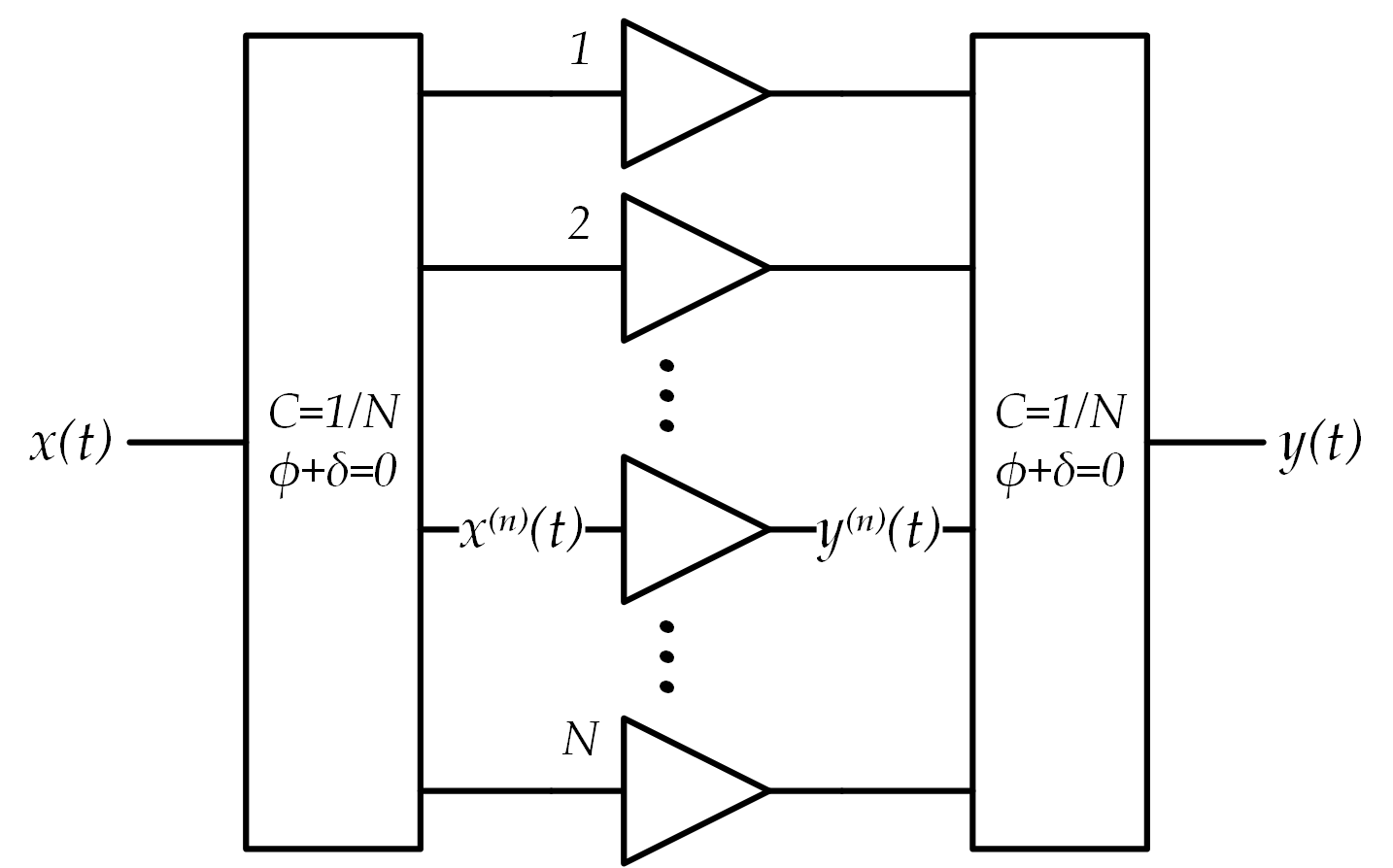}}
	\caption{A parallel combination of balanced amplifiers.}
	\label{fig:Model_N}
\end{figure}

A set of $N$ balanced amplifiers may be combined using power splitters/combiners (e.g. hybrids with $\phi=0$), as shown in Fig.~\ref{fig:Model_N}.
For a $N$-way splitter, there will be $N$ balanced amplifiers, and thus $2N$ total, single-ended amplifiers.
The input signal to each balanced amplifier is then 
\begin{equation}
	x^{(n)}(t) = \frac{1}{\sqrt{N}} A \left[ \cos \left( \omega_1 t \right) + \cos \left( \omega_2 t \right) \right]
\end{equation}
Each amplifier then outputs 
\begin{equation}
\begin{split}
	y^{(n)}(t) & = a_1 \frac{1}{\sqrt{N}} A \cos \left( \omega_1 t \right) + \frac{3}{4} a_3  \left( \frac{1}{\sqrt{N}} A \right)^3 \cos \left( (2\omega_1 - \omega_2) t \right)  + \cdots
\end{split}
\end{equation}
The output of the parallel amplifier is then
\begin{equation}
	y(t) = a_1 A \cos \left( \omega_1 t \right) + \frac{3}{4} a_3  \frac{1}{N} A^3 \cos \left( (2\omega_1 - \omega_2) t \right)  + \cdots
\end{equation}
From Eq.~(\ref{eqn:BallancedIIP}) we find the IIP3 for the parallel amplifier in Fig.~\ref{fig:Model_N} is 
\begin{equation}
	A_{IIP3}^{(N)} = \sqrt{N} A_{IIP3}^{(1)}
\end{equation}
In other words, arranging $N$ balanced amplifiers in parallel improves IIP3 by a factor of $N$ in power.

\subsection{Sensitivity}

Assuming lossless hybrids, the noise power output from a balanced amplifier is equal to the noise power output from one of the constituent amplifiers when the input is matched.
This 
follows from the fact that (a) the noise from the constituent amplifiers is uncorrelated~\cite{Kerr98}, and (b) only half the noise power from each amplifier appears at the output, the other half being absorbed by the hybrid's terminated port (see e.g.~\cite{Kurokawa1965}).
Therefore $T_R$ does not increase in a balanced configuration
relative to the NF of the constituent single-ended amplifier, as long as the hybrids are lossless.

\begin{figure}
	\centering{\includegraphics[scale=1]{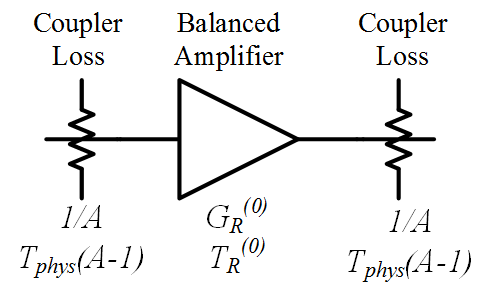}}
	\caption{Noise model for a combined amplifier, in which the loss internal to the hybrids has been modeled as external attenuation to a combined amplifier with lossless hybrids.}
	\label{fig:NoiseModel}
\end{figure}

Since practical hybrids are not perfectly lossless, the hybrids degrade the NF.
If $A$ is the attenuation (reciprocal gain) of the hybrids, as in Fig.~\ref{fig:NoiseModel}, the input-referred noise temperature for the cascade hybrid-amplifier-hybrid is
\begin{equation}
\begin{split}
	T_R & = T_{phys}(A-1) + \frac{T_R^{(0)}}{1/A} + \frac{T_{phys}(A-1)}{G_R^{(0)}/A} \\
\end{split}
	\label{T_p1} 
\end{equation}
where $T_{phys}$ is the physical temperature of the hybrids and $T_R^{(0)}$ and $G_R^{(0)} = |a_1|^2$ are the constituent amplifier's noise temperature and gain, respectively.
In the special case of a lossless hybrids, $A=1$ and subsequently $T_R = T_R^{(0)}$, as expected.

\section{Design Examples}
\label{sec:DesignExamples}
This section discusses the design and testing of two front ends to demonstrate the findings of the previous sections.
A high-sensitivity front end based on the PGA-103 is presented in Section~\ref{sec:PGA}, as well front ends which combine multiple PGA-103s in series and parallel configurations.
A high-linearity front end based on the HELA-10 is presented in Section~\ref{sec:HELA}.

\subsection{PGA-103 Based Front Ends in Various Configurations}
\label{sec:PGA}
\begin{figure}
	\centering
	\includegraphics[scale=0.75]{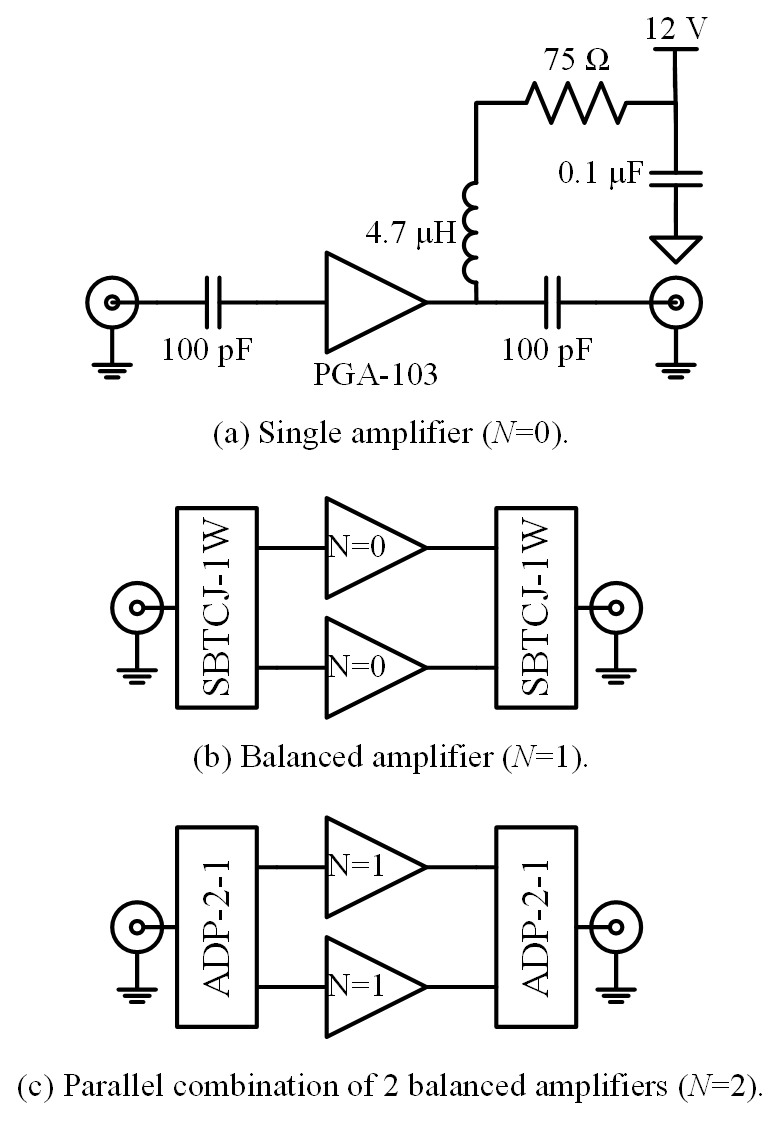}
	\caption{Schematics of the single, balanced, and parallel combined amplifiers tested in Section~\ref{sec:PGA}.}
	\label{fig:Sch}
\end{figure}

\begin{figure}
	\centering
\begin{center}
\begin{tabular}{ccc}
\includegraphics[width=1.6in]{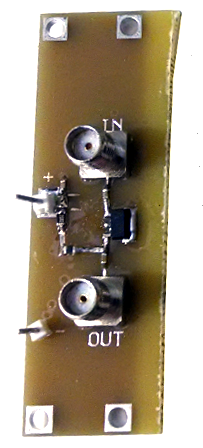} & 
\includegraphics[width=1.7in]{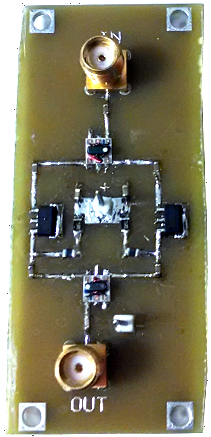} & 
\includegraphics[width=3.0in]{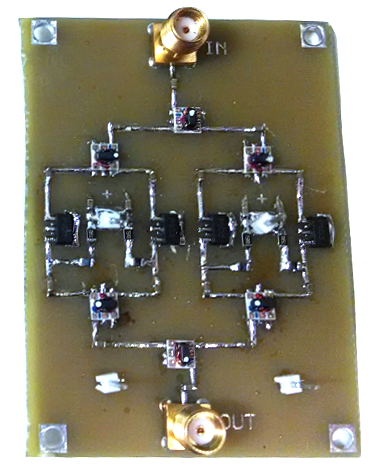}\\
\end{tabular}
\end{center}
	\caption{Fabricated (\textit{left to right}) single, balanced, and parallel combined amplifiers tested in Section~\ref{sec:PGA}.}
	\label{fig:Brd}
\end{figure}

The MC PGA-103 amplifier was identified in the survey described in Section~\ref{sec:Survey} as exceptional with respect to comparable devices, particularly in its sensitivity (see Fig.~\ref{fig:Survey}).  
This device achieves a gain of +26.8~dB with an IIP3 of 10.4~dBm and $T_R$=36~K at 50~MHz when biased at 90~mA.
Single-amplifier ($N=0$), balanced ($N=1$), and parallel balanced ($N=2$) versions of PGA-103-based amplifiers were constructed and tested.  
Schematic diagrams and the tested units are shown in Figs.~\ref{fig:Sch}~and~\ref{fig:Brd}, respectively.
The $N=1$ design uses the MC SBTCJ-1W+ ($\phi=180^\circ$, $\delta=0.4^\circ$ typical, $A$=0.6~dB at 50~MHz). 
The $N=2$ design is a parallel combination of two $N=1$ devices using MC ADP-2-1 ($\phi=0^\circ$, $\delta=0.2^\circ$ typical, $A$=0.2~dB at 50~MHz) power splitter/combiners. 
Current provided to each amplifier was within 5\% of 90~mA.

\begin{figure}
	\centering
	\includegraphics[width=\textwidth]{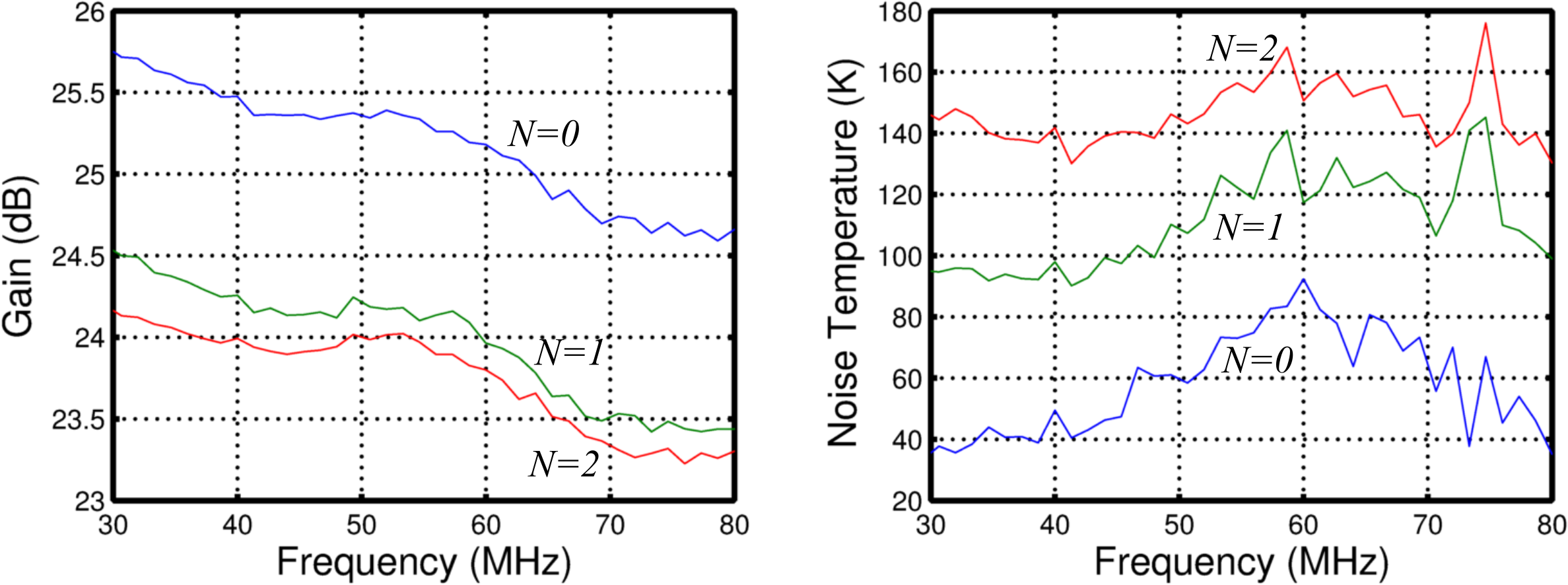}
	\caption{ Measured gain and noise temperature of the PGA-103 front ends.} 
	\label{fig:JAI_PGA_mod}
\end{figure}

Figure~\ref{fig:JAI_PGA_mod} shows the measured gain and noise temperature for each PGA-103 front end.
As expected, gain decreases slightly for $N=1$ with respect to $N=0$, and $N=2$ with respect to $N=1$, due to losses in the hybrids.
$T_R$ increases for the same reason, as does the power consumption, from 1.1~W for the $N=0$ front end to 2.2~W and 4.3~W for the $N=1$ and $=2$ front ends, respectively. 

\begin{figure}
	\centering{\includegraphics[scale=0.8]{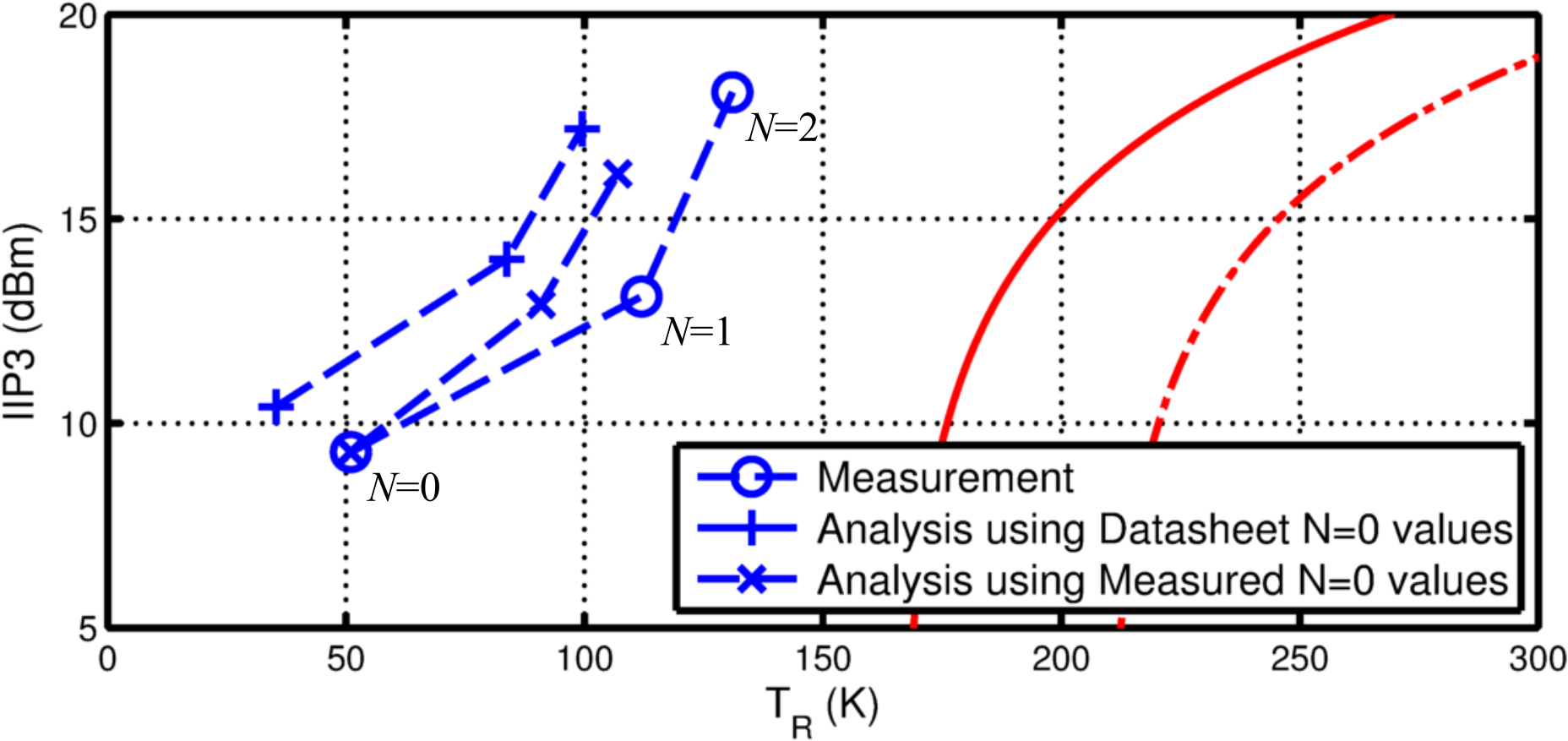}}
	\caption{Comparison of measurements for PGA-103 based front ends from $N=0$ to $N=2$.  Also shown are the results for $N=0$ through $N=2$ predicted from the analysis using only data for the $N=0$ device, obtained from the data sheet.
		The dot-dash and solid lines are the same curves appearing in Fig.~\ref{fig:Survey}, provided as a reference.}
	\label{fig:Model}
\end{figure}

Figure~\ref{fig:Model} summarizes the test results at 50~MHz in the form of the analysis of Section~\ref{sec:Survey}.
For comparison, theoretical results are shown assuming both the nominal (datasheet) and measured $N=0$ amplifier performance, following the analysis of the previous section and accounting for the hybrids' loss and phase errors.  
The measured $T_R$ of the $N=0$ design was within 15~K of the expected value, and the measured IIP3 was within 1~dB of the expected value.
Errors can reasonably be attributed to a combination of component manufacturing variances and limitations of the test methods and equipment. 

The $N=1$ design improves IIP3 by 3.8~dB and degrades $T_R$ by 61~K; this compared to 3.6~dB and 50~K, respectively, predicted by the analysis 
using datasheet values from the $N=0$ device.
The $N=2$ design improves IIP3 by an additional 5~dB and degrades $T_R$ by an additional 19~K; this compared to predicted values of 3.2~dB and 16~K, respectively, again using datasheet values for the PGA-103 and SBTCJ-1W.  
These results are consistent with the analysis presented in Section~\ref{sec:AmpModel}.

For the $N=0$ amplifier, IIP2 was measured to be $+$21~dBm (no corresponding value is provided by the manufacturer).
IIP2 for the $N=1$ device is predicted to increase by almost 60~dB relative to the $N=0$ device, to nearly $+$80~dBm.  
The achieved value was beyond our ability to measure (i.e. greater than $+$70~dBm). 

\subsection{A HELA-10-based Front End for Strong Interference Environments}
\label{sec:HELA}
The MC HELA-10 is the most linear MMIC amplifier identified in the survey described in Section~\ref{sec:Survey}. 
Here, we describe a new HELA-10 based front end for use between 30-80~MHz at sites at where increased linearity is required to accommodate strong radio frequency interference (RFI).

\begin{figure}
	\centering
	\includegraphics[scale=0.85]{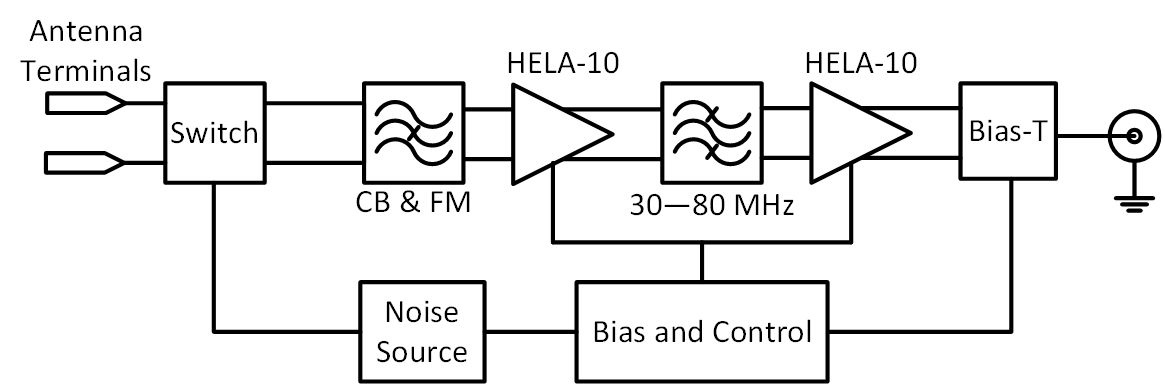}
	\caption{Block diagram of the HELA-10 based front end.}
	\label{fig:FEE_bd}
\end{figure}

\begin{figure}
	\centering
	\includegraphics[width=\textwidth]{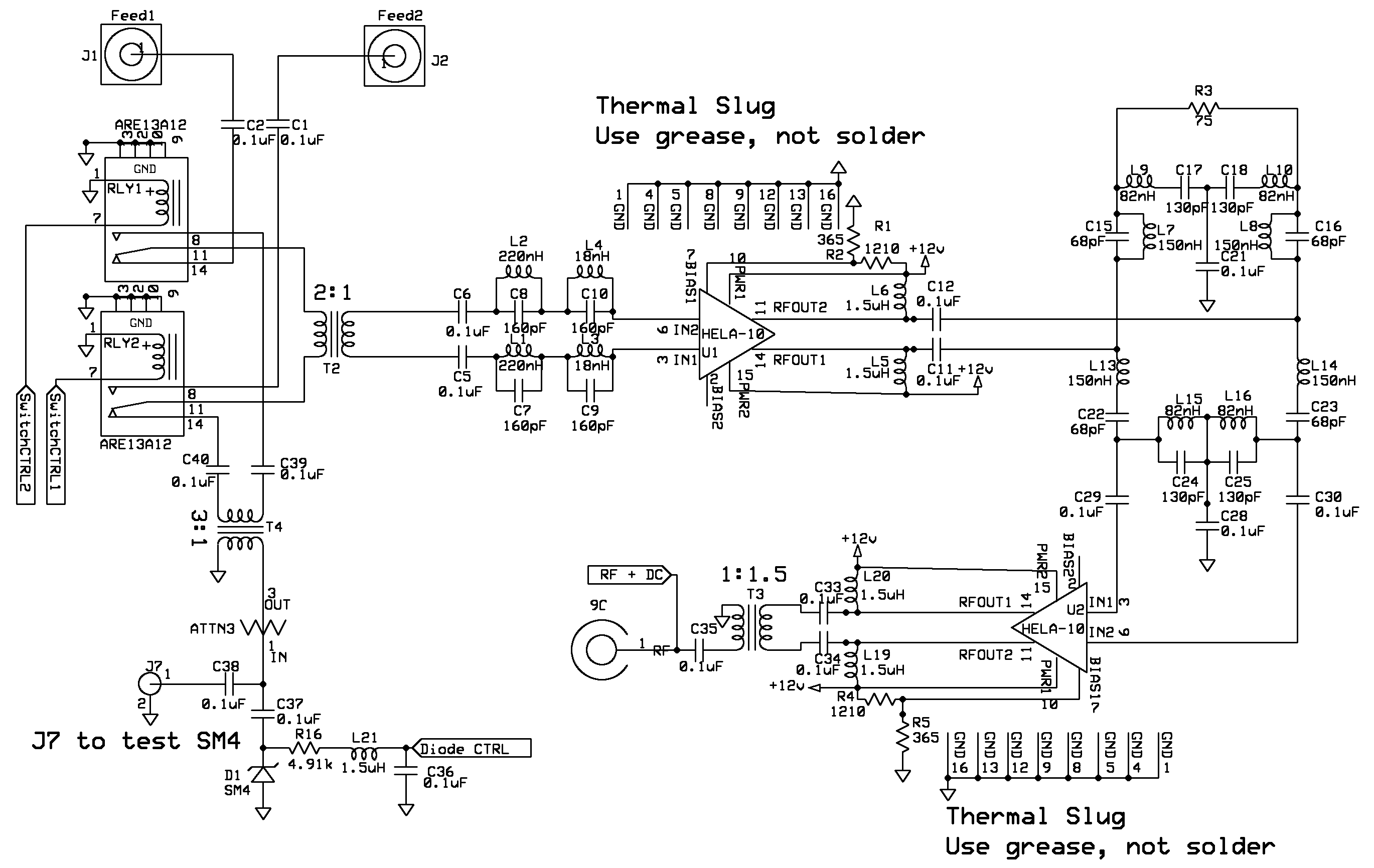}
	\caption{Schematic of the HELA-10 based front end.
		``Feed1'' and ``Feed2'' are the antenna terminals.}
	\label{fig:FEE_sch}
\end{figure}

\begin{figure}
	\centering
	\includegraphics[width=\textwidth]{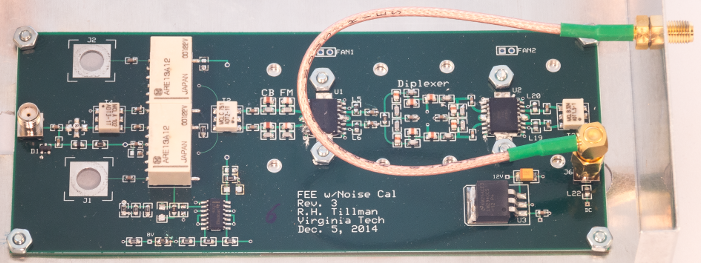}
	\caption{The HELA-10 based front end.}
	\label{fig:FEE_built}
\end{figure}

A block diagram and schematic of the front end are shown in Figs.~\ref{fig:FEE_bd}~and~\ref{fig:FEE_sch}, respectively.
The two HELA-10 amplifiers are used as a pre-amplifier and a line driver, respectively.
Preceding the first amplifier are notch filters designed to trap the strong signals from Citizen Band (CB) and FM radio at 26.5--27.5 and 88--108 MHz, respectively.
An absorptive second-order Butterworth filter sets the bandpass of the front end to 30-80~MHz. 
The signal path is differential from the antenna through the output of the front end, at which point a balun converts the signal path to $50~\Omega$ single-ended.
The front end draws $\sim$0.8~A of DC current at a DC voltage ranging from 16-20~V.
Additionally, the front end implements a three-state calibration system similar to that used by the Large aperture Experiment to detect the Dark Ages (LEDA)~\cite{Greenhill_LEDA_2012} and Experiment to Detect the Global EoR Step (EDGES)~\cite{RogersBowman_2012} to facilitate absolute measurement of $T_A$.
The calibration state is determined by the DC voltage used to bias the front end~\cite{TillmanEllingson2015a}.
Figure~\ref{fig:FEE_built} shows a completed front end, affixed to its enclosure, which doubles as a heat sink.

\begin{figure}
	\centering
	\includegraphics[width=\textwidth]{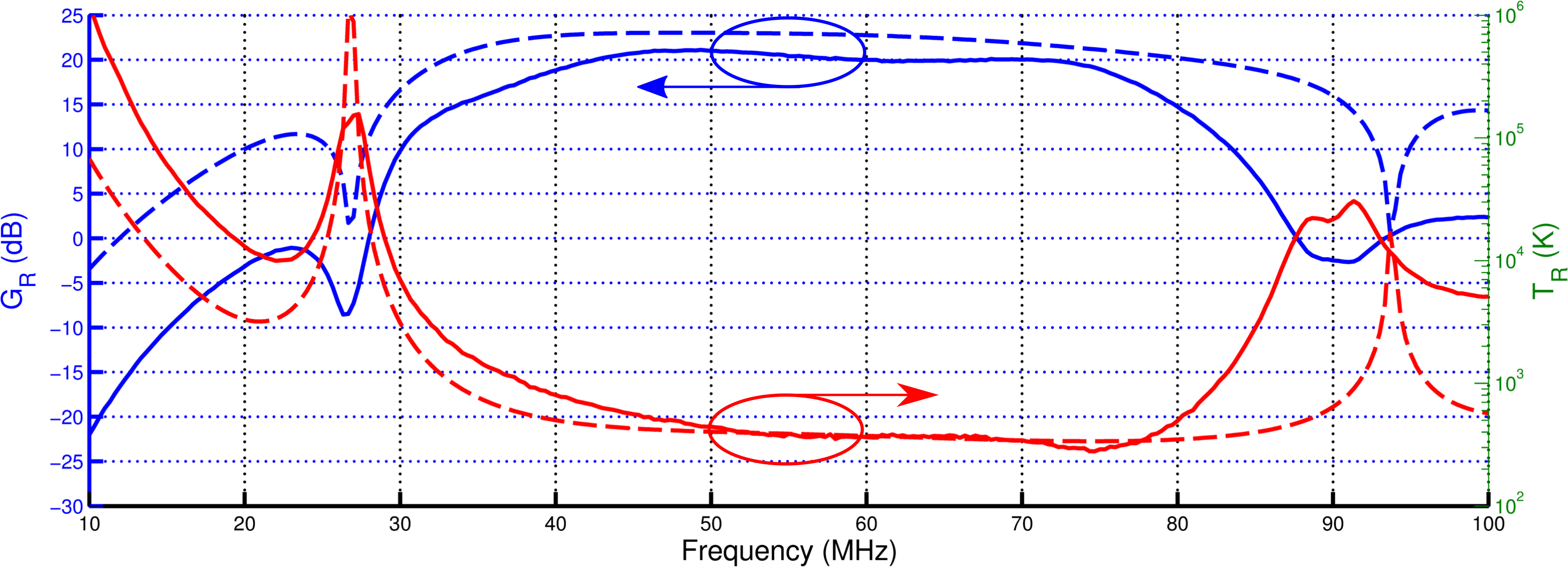}
	\caption{({\it Solid:}) Measured and ({\it dashed:}) predicted gain and noise temperature of the HELA-10 front end.} 
	\label{fig:LabCommisioning_JAI_mod}
\end{figure}

Figure~\ref{fig:LabCommisioning_JAI_mod} shows the typical (average from the four constructed units) front end gain and noise temperature.
The measured $P_{1dB}$ and IIP3 are $+$11~dBm and $+$24.9~dBm, respectively, at 50~MHz, and the achieved IIP2 was again beyond our ability to measure.

\section{Field Demonstration}
\label{sec:Demo}
The goal of the demonstration is to accurately measure the Galactic noise background.
Equation~\ref{eqn:lfa_eTsky} approximates the contribution of this background noise to the antenna temperature.

\begin{figure}
	\centering
	\includegraphics[scale=0.85]{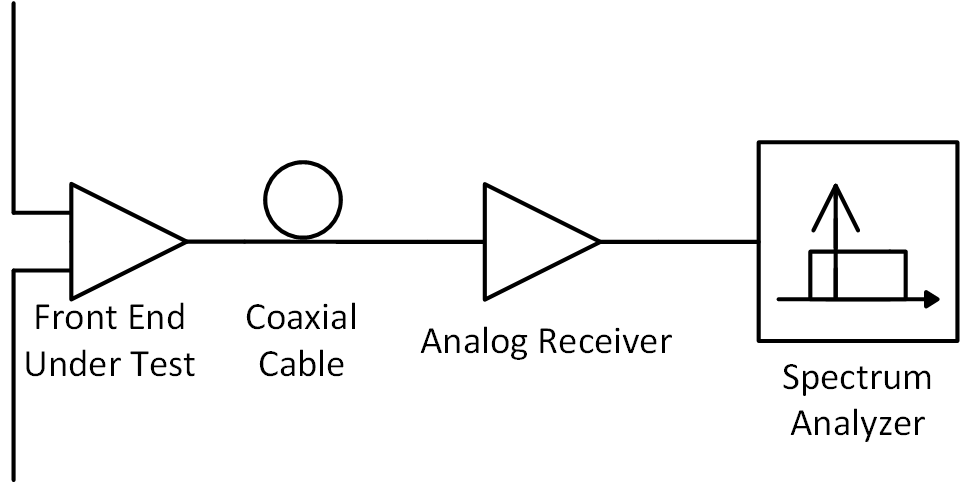}
	\caption{Block diagram of the test setup for the demonstration presented in Section~\ref{sec:Demo}.}
	\label{fig:FieldTest}
\end{figure}

A block diagram of the demonstration setup is shown in Fig.~\ref{fig:FieldTest}.
The antenna used in this demonstration is a 3.048 m long, 2.5 cm diameter, straight copper dipole positioned parallel to and 1.524 m above a 7.6~m$\times$7.6~m ground screen, which sits on flat, grass-covered ground.
The coaxial cable is 152~m of LMR-400.
The analog receiver provides a gain of about 75~dB, 15-73~MHz bandwidth (3~dB), and input-referred noise temperature of $720$~K.
The coaxial cable, analog receiver, and spectrum analyzer are calibrated out and do not affect the results.

\begin{figure}
	\centering
	\includegraphics[width=\textwidth]{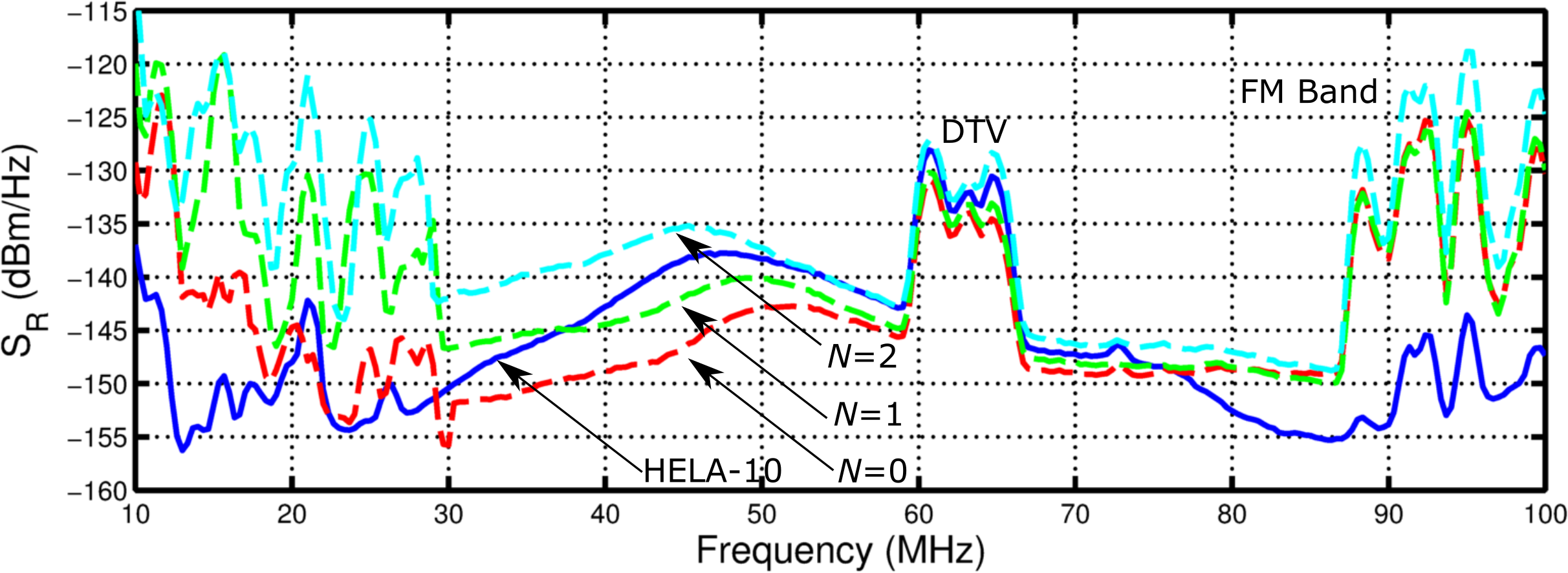}
	\caption{ Measured PSD for each front end, referenced to the front end output (input to the coaxial cable). See text for discussion of continuum power available from the antenna for these experiments.
		1~MHz resolution bandwidth, 33~ms integration.}
	\label{fig:JAI_SkySpectra_S_R}
\end{figure}

All measurements were taken in Glen Alton, VA; a rural area about 65 km northwest of Blacksburg, VA.   All measurements were taken at approximately the same local sidereal time, within one hour.
Figure~\ref{fig:JAI_SkySpectra_S_R} shows the measured PSD for each front end.
The expected HF-band RFI is present below 30~MHz, as well as the FM band between 87-108~MHz.
The large signal between 60-66~MHz is a digital television (DTV) station (Channel~3).
The total power available from the antenna (i.e., applied to the FEE input) in the spectrum shown is approximately $-$22~dBm. 
However the signal available from the antenna is strongly dominated by local AM stations transmitting at the 5-10 kW level, resulting in an actual total power of at least $-$10 dBm from our antenna.  
The power associated with the AM stations was not accurately measured at the time of the experiment and may have actually been significantly greater.

\begin{figure}
	\centering
	\includegraphics[scale=0.75]{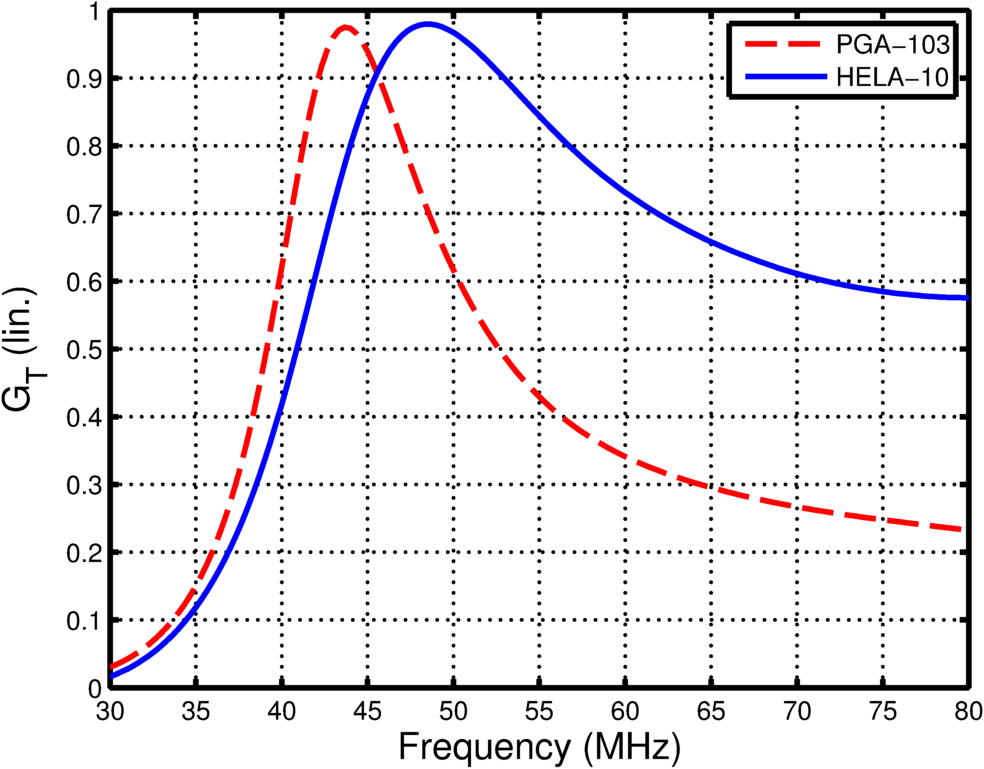}
	\caption{Estimated $G_T$ of the antenna interface for the field measurements in Sec.~\ref{sec:Demo}.}
	\label{fig:JAI_SkySpectra_G_T}
\end{figure}

The antenna temperature for each front end was estimated by solving Eq.~(\ref{eqn:Sr}) for $T_{A}$ as follows:
\begin{equation}
	T_{A} = \frac{S_R}{k G_T G_R} - \frac{T_R}{G_T} 
	\label{eqn:T_sky_meas}
\end{equation}
%
For the PGA-103 front ends, $G_R$ and $T_R$ where measured in the lab.
For the HELA-10 front end, $G_R$ and $T_R$ where determined using the internal calibration system.
The single-ended PGA-103 front ends were connected to the antenna through an MC ADT1-1 balun transformer, which has a 1:1 impedance ratio and a measured insertion loss of about 0.5~dB over the considered frequency range.
The antenna impedance was estimated using Ansys HFSS EM modeling software, and Fig.~\ref{fig:JAI_SkySpectra_G_T} shows the resulting $G_T$.
The antenna temperature is related to the temperature distribution surrounding the antenna $T(\theta,\phi)$ by~\cite{Stutzman1998}
\begin{equation}
	T_A = \eta \left[ \frac{1}{\Omega_A} \int_{0}^{\pi} \int_{0}^{2\pi} T(\theta,\phi) P(\theta,\phi) d\Omega \right]
\end{equation}
where $P(\theta,\phi)$ and $\Omega_A$ are the antenna's normalized power pattern and beam solid angle, respectively, and $d\Omega = \sin \theta d\phi d\theta$ is the element of solid angle.
The contributions of $T_{sky}$ and $T_{gnd}$ to $T_A$ are the upper and lower hemisphere portions of the $T_A$ integral:
\begin{equation}
	T_{sky} = \frac{1}{\Omega_A} \int_{0}^{\pi/2} \int_{0}^{2\pi} T(\theta,\phi) P(\theta,\phi) d\Omega
	\quad , \quad
	T_{gnd} = \frac{1}{\Omega_A} \int_{\pi/2}^{\pi} \int_{0}^{2\pi} T(\theta,\phi) P(\theta,\phi) d\Omega
\end{equation}
where we have assumed ground loss to be negligible, such that $\eta=1$ and the ground perfectly reflects the sky temperature distribution.
Using a moment method model (NEC2, assuming a ground conductivity of 5~mS/m and dielectric constant of 15) to calculate $P(\theta,\phi)$ and assuming a uniform brightness temperature distribution with a constant temperature given by Eq.~(\ref{eqn:lfa_eTsky}), we find $T_{gnd} \approx 0.045 T_{sky}$ across the entire frequency range of interest.
Thus, the sky temperature was estimated by $T_{sky} = T_A/1.045$. 

\begin{figure}
	\centering
	\includegraphics[width=\textwidth]{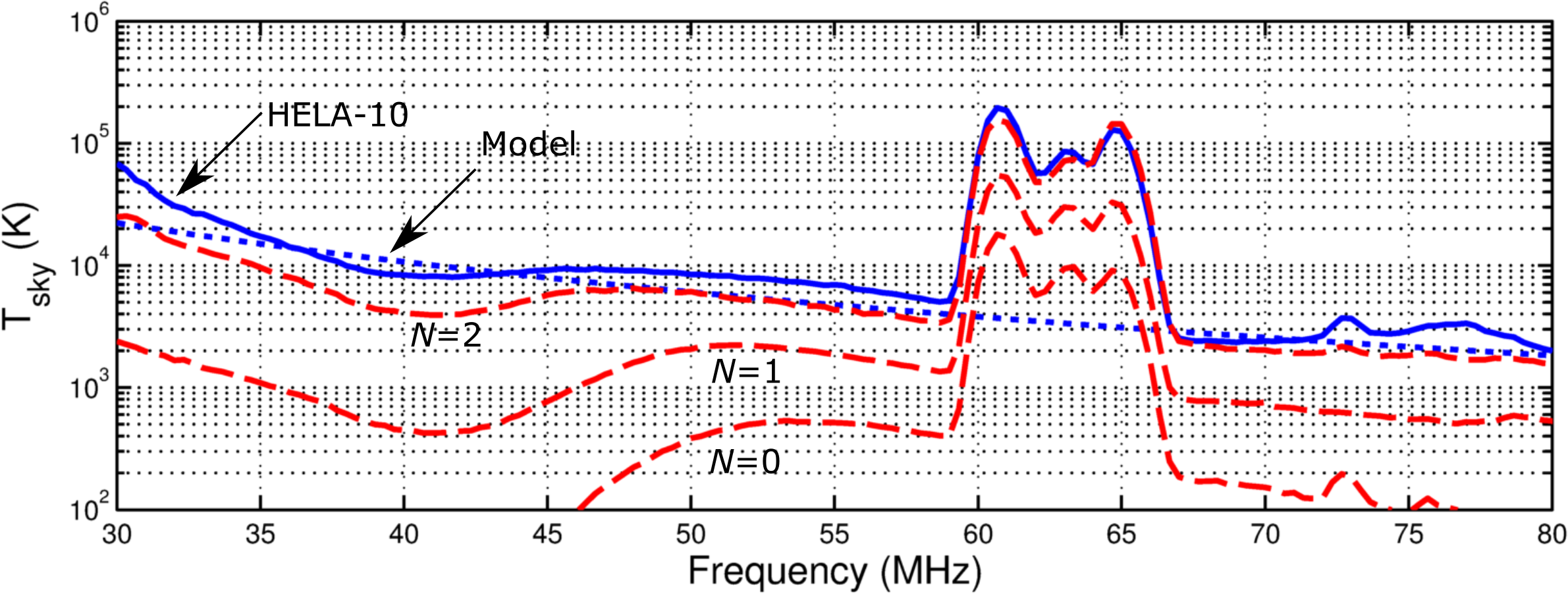}
	\caption{ ({\it Dashed:}) Estimated $T_{sky}$ using the PGA-103 Front Ends.
		({\it Solid:}) Estimated $T_{sky}$ using the HELA-10 Front End.
		({\it  Dotted:}) Model of $T_{sky}$ given in Eq.~(\ref{eqn:lfa_eTsky}).}
	\label{fig:JAI_SkySpectra_cal}
\end{figure}

Figure~\ref{fig:JAI_SkySpectra_cal} shows $T_{sky}$ estimated using Eqs.~(\ref{eqn:TA})~and~(\ref{eqn:T_sky_meas}) and the data collected from each front end, and the modeled $T_{sky}$ from Eq.~(\ref{eqn:lfa_eTsky}).
The HELA-10 front end is not noticeably affected by the interference, due in part to the highly linear HELA-10 amplifiers and in part to the CB, FM, and bandpass filters.
Remaining disagreement with the model is attributed to uncertainty in the antenna impedance.

Although the $N=0$ PGA-103 front end appears to be experiencing severe compression, we are not certain if this is classical gain compression, or whether it is a more complex non-linear behavior arising from the spectral distribution of the interference (i.e., the AM-dominated antenna output power) or some other effect resulting from the power level handled by a single PGA-103 in this case.
The improved linearity of the balanced ($N=1$) and parallel ($N=2$) PGA-103 front is apparent, as the estimated $T_{sky}$ approaches the model for increasing $N$.
Without the interference, or with the addition of input filters similar to those on the HELA-10 front end, we expect the PGA-103 front ends would yield a $T_{sky}$ similar to the result from the HELA-10 front end.

\section{Conclusions}
\label{sec:conc}
This paper has considered the design of front ends for modern low frequency radio telescopes with respect to the trade off between sensitivity and linearity.
A survey of commercially-available MMIC amplifiers was performed and an informal analysis was used to identify the state-of-the-art.
Two outlying amplifiers, which outperformed the majority of comparable amplifiers, were selected for demonstration.
A high sensitivity front end using the PGA-103 amplifier was presented with $T_R \approx 50$~K, however this amplifier was significantly affected by interference in a field demonstration.
A high-linearity front end using HELA-10 amplifiers and CB/FM-band filters was less sensitive than the PGA-103 front end ($T_R \approx 710$~K), but was robust to the interference and facilitated a greatly improved detection of the Galactic synchrotron background.

Balanced and parallel configurations of amplifiers were also considered with respect to the linearity-sensitivity trade off.
Front ends employing PGA-103 amplifiers in balanced and parallel configurations reduced the effects of non-linearity, as expected from theory, at a cost to both sensitivity and power consumption.


\section*{Acknowledgments}
This work was supported in part by the Virginia Tech Bradley Foundation and in part by the National Science Foundation through Grants AST-1105949 and AST-1139963. 
The authors acknowledge the assistance of X.~Gomez of Virginia Tech, for providing the HFSS antenna impedance model and for assisting with the field measurement.

\appendix{Intercept Points of a Balanced Amplifier}
\label{sec:App_Der}
For a signal $x(t) = A [ \cos \omega_1 t + \cos \omega_2 t ]$ input to a balanced amplifier as in Fig.~\ref{fig:QuadComb}, the inputs to each amplifier are
\begin{subequations}
\begin{equation}
	x^{(A)}(t) = \frac{1}{\sqrt{C}} A \left[ \cos \left( \omega_1 t \right) + \cos \left( \omega_2 t \right) \right]
\end{equation}
\begin{equation}
	x^{(B)}(t) = \frac{1}{\sqrt{C}} A \left[ \cos \left( \omega_1 t  - \phi  - \delta \right) + \cos \left( \omega_2 t - \phi  - \delta \right) \right]
\end{equation}
\end{subequations}
The outputs from the amplifiers are then 
\begin{subequations}
\begin{equation}
\begin{split}
	y^{(A)}(t) & = a_1 \frac{1}{\sqrt{C}} A \left[ \cos \left( \omega_1 t \right) + \cos \left( \omega_2 t \right) \right] \\
		& \quad + a_2 \left( \frac{1}{\sqrt{C}} A \right)^2 \left[ \cos \left( (\omega_1 - \omega_2) t \right) + \cos \left( (\omega_2 + \omega_1) t \right) \right] \\
		& \quad + \frac{3}{4} a_3 \left( \frac{1}{\sqrt{C}} A \right)^3 \left[ \cos \left( (2\omega_1 - \omega_2) t \right) + \cos \left( (2\omega_1 + \omega_2) t \right) \right. \\
		& \quad ~~~~~~~~~~~~~~~~~~~~~~ \left. + \cos \left( (2\omega_2 - \omega_1) t \right) + \cos \left( (2 \omega_2 + \omega_1) t \right) \right]
\end{split}
\end{equation}
and
\begin{equation}
\begin{split}
	y^{(B)}(t) & = a_1 \frac{1}{\sqrt{C}} A \left[ \cos \left( \omega_1 t - \phi - \delta \right) + \cos \left( \omega_2 t - \phi - \delta \right) \right] \\
		& \quad + a_2 \left( \frac{1}{\sqrt{C}} A \right)^2 \left[ \cos \left( (\omega_1 - \omega_2) t \right) + \cos \left( (\omega_2 + \omega_1) t  - 2\phi - 2\delta\right) \right] \\
		& \quad + \frac{3}{4} a_3 \left( \frac{1}{\sqrt{C}} A \right)^3 \left[ \cos \left( (2\omega_1 - \omega_2) t - \phi - \delta \right) + \cos \left( (2\omega_1 + \omega_2) t - \phi - \delta \right) \right. \\
		& \quad ~~~~~~~~~~~~~~~~~~~~~~ \left. + \cos \left( (2\omega_2 - \omega_1) t - \phi - \delta \right) + \cos \left( (2 \omega_2 + \omega_1) t - \phi - \delta \right) \right]
\end{split}	
\end{equation}
\end{subequations}
respectively, where the DC, harmonic distortion, and gain compression terms have been suppressed (e.g. assuming filtering) for the purposes of determining IIP2 and IIP3.
The output hybrid modifies each term in $y^A(t)$ by a phase shift of $\phi + \delta$ and adds it to $y^B(t)$, such that
\begin{equation}
\begin{split}
	y(t)  & = a_1 \frac{2}{C} A \left[ \cos \left( \omega_1 t - \phi - \delta \right) + \cos \left( \omega_2 t - \phi - \delta \right) \right] \\
		& \quad + a_2 \frac{1}{\sqrt{C}} \left( \frac{1}{\sqrt{C}} A \right)^2 \left[ \cos \left( (\omega_1 - \omega_2) t - \phi - \delta \right) + \cos \left( (\omega_2 + \omega_1) t - \phi - \delta \right) \right. \\
		& \quad ~~~~~~~~~~~~~~~~~~~~~~~~~ \left. + \cos \left( (\omega_1 - \omega_2) t \right) + \cos \left( (\omega_2 + \omega_1) t  - 2\phi - 2\delta\right) \right] \\
		& \quad + \frac{3}{4} a_3 \frac{2}{\sqrt{C}} \left( \frac{1}{\sqrt{C}} A \right)^3 \left[ \cos \left( (2\omega_1 - \omega_2) t - \phi - \delta \right) + \cos \left( (2\omega_1 + \omega_2) t - \phi - \delta \right) \right. \\
		& \quad ~~~~~~~~~~~~~~~~~~~~~~~~~~~ \left. + \cos \left( (2\omega_2 - \omega_1) t - \phi - \delta \right) + \cos \left( (2 \omega_2 + \omega_1) t - \phi - \delta \right) \right]
\end{split}
\end{equation}
For a $180^\circ$ hybrid, $\phi=\pi$, and for commercially-available hybrids operating at VHF and UHF $\delta$ is typically less than $5^\circ$.
Expanding the trigonometric functions in order to isolate $\delta$, applying small angle approximations for $\delta$, and keeping only the first-order $\delta$ terms, the output becomes
\begin{equation}
\begin{split}
	y(t)  & \cong - a_1 \frac{2}{C} A \left[ \cos(\omega_1 t)  + \cos (\omega_2 t) \right. \\
		& \quad ~~~~~~~~~~~~~~~~ \left. + \delta \sin(\omega_1 t)  + \delta \sin (\omega_2 t) \right] \\
		& \quad - \delta a_2 \frac{1}{\sqrt{C}} \left( \frac{1}{\sqrt{C}} A \right)^2 \left[  \sin \left( (\omega_1 - \omega_2) t  \right) - \sin \left( (\omega_2 + \omega_1) t \right) \right] \\
		& \quad - \frac{3}{4} a_3 \frac{2}{\sqrt{C}} \left( \frac{1}{\sqrt{C}} A \right)^3 \left[ \cos \left( (2\omega_1 - \omega_2) t \right) + \cos \left( (2\omega_1 + \omega_2) t \right) \right. \\
		& \quad ~~~~~~~~~~~~~~~~~~~~~~~~~~~ \left. + \cos \left( (2\omega_2 - \omega_1) t \right) \right. + \cos \left( (2 \omega_2 + \omega_1) t \right) \\
		& \quad ~~~~~~~~~~~~~~~~~~~~~~~~~~~ + \delta \sin \left( (2\omega_1 - \omega_2) t \right) + \delta \sin \left( (2\omega_1 + \omega_2) t \right)  \\
		& \quad ~~~~~~~~~~~~~~~~~~~~~~~~~~~ + \delta \sin \left( (2\omega_2 - \omega_1) t \right) \left. + \delta \sin \left( (2 \omega_2 + \omega_1) t \right) \right]
\end{split}
\end{equation}
of which
\begin{equation}
	y_{0,i} = - a_1 \frac{2}{C} A ( \cos(\omega_i t) + \delta \sin (\omega_i t) \quad , \quad i=1,2
\end{equation}
are the linear, fundamental tones,
\begin{equation}
	y_{IM2,ij} = \delta a_2 \frac{1}{\sqrt{C}} \left(\frac{1}{\sqrt{C}} A \right)^2 \sin \left( (\omega_i + \omega_j) t \right) \quad , \quad i,j=1,2, j \neq i
\end{equation}
are the IM2 products, and
\begin{equation}
	y_{IM3,ij} = - \frac{3}{4} a_3 \frac{2}{\sqrt{C}} \left( \frac{1}{\sqrt{C}} A \right)^3  \left[ \cos \left( (2\omega_i - \omega_j) t \right) + \delta \sin \left( 2(\omega_i - \omega_j) t \right) \right] \quad , \quad i,j=1,2, j \neq i
\end{equation}
are the IM3 products.
To determine IIP2 and IIP3, we need the root-mean-square (RMS) values of the fundamental tones and the intermodulation products.
The mean is taken by averaging over each tone's period $T$ (denoted by $\langle \cdot \rangle$), and any terms containing $\delta^2$ are considered negligible.
The result is
\begin{subequations}
\begin{equation}
\begin{split}
	\sqrt{ \left\langle y_0^2(t)\right\rangle } \cong \frac{1}{\sqrt{2}}  a_1 \frac{2}{C} A 
\end{split}
\end{equation}
\begin{equation}
\begin{split}
	\sqrt{ \left\langle y_{IM2}^2(t)\right\rangle }  \cong \frac{1}{\sqrt{2}} \delta a_2 \frac{1}{\sqrt{C}} \left( \frac{1}{\sqrt{C}} A \right)^2 
\end{split}
\end{equation}
\begin{equation}
\begin{split}
	\sqrt{ \left\langle y_{IM3}^2(t)\right\rangle } \cong \frac{3}{4 \sqrt{2}} a_3 \frac{2}{\sqrt{C}} \left( \frac{1}{\sqrt{C}} A \right)^3
\end{split}
\end{equation}
\label{eqn:rms}
\end{subequations}
Finally, from (\ref{eqn:N0}) and Eq.~(\ref{eqn:rms}), the IIP2 and IIP3 for the single pair of balanced amplifiers are found to be
\begin{equation}
	A_{IIP2}^{(1)} \cong 2 \frac{\sqrt{C}}{\delta} A_{IIP2}^{(0)} \quad , \quad A_{IIP3}^{(1)} \cong \sqrt{C} A_{IIP3}^{(0)}
\end{equation}

\end{document}